# Light transport in $\mathcal{PT}$-invariant photonic structures with hidden symmetries

M.H. Teimourpour and R. El-Ganainy[*]
*Department of Physics, Michigan Technological University, Houghton, Michigan 49931, USA*

A. Eisfeld
*Max Planck Institute for the Physics of Complex systems, Nothnitzer Street 38, Dresden, Germany*

A. Szameit
*Institute of Applied Physics, Abbe School of Photonics, Friedrich-Schiller-University, Max-Wien-Platz 1, D-07743 Jena, Germany*

D.N Christodoulides
*College of Optics & Photonics-CREOL, University of Central Florida, Orlando, Florida, USA*

**Abstract**

We introduce a recursive bosonic quantization technique for generating classical PT photonic structures that possess hidden symmetries and higher order exceptional points. We study light transport in these geometries and we demonstrate that perfect state transfer is possible only for certain initial conditions. Moreover, we show that for the same propagation direction, left and right coherent transports are not symmetric with field amplitudes following two different trajectories. A general scheme for identifying the conservation laws in such PT-symmetric photonic networks is also presented.

[*]ganainy@mtu.edu

Parity-time (PT) symmetry in quantum mechanics has been a subject of intense investigations in the past decade. This interest was sparked by the seminal work of Bender et al [1] in which they demonstrated that a certain class of PT-symmetric Hamiltonians can exhibit entirely real eigenvalue spectra. Within the context of single particle Schrodinger equation, PT symmetry arises from complex potentials having even/odd symmetry for their real/imaginary components [1-4]. Above a specific threshold for the imaginary part of the complex potential, the system experiences an abrupt phase transition where some real eigenvalues branch over to the complex plane, signaling the onset of spontaneous PT symmetry-breaking [1-4]. These phase transition thresholds, also known as exceptional points [5,6], are marked by the coalescing of two or more eigenvectors.

The notion of PT-symmetry was later extended to optical waveguide systems with complex refractive index profiles [7-9]. Similar to their quantum mechanical counterparts, the real/imaginary components of the optical indices in these arrangements are even and odd functions, respectively [7-12]. PT symmetry in complex Bragg gratings [13] and optical cavities [14] have been also studied. In addition, non-Hermitian optical diodes have been proposed [15]. Within the context of coupled optical structures, PT symmetry has been mainly investigated in uniform lattices [8-10]. Nevertheless, very recently $J_x$ arrays [16] exhibiting PT symmetry have been considered and their spectra were calculated using angular momentum operator algebra [17]. The dynamical properties of these PT coupled structures were also investigated and it was shown that they exhibit half cycle conditional perfect state transfer (PST) below PT phase transition [18].

In this work we introduce recursive bosonic quantization (RBQ) technique for generating new classes of PT symmetric networks that exhibit nonlinear topology and possess hidden symmetry. The latter property gives rise to spectral degeneracies that persist even above PT spontaneous symmetry breaking. We also investigate light transport in these geometries and we demonstrate that perfect state transfer is possible only for certain initial conditions. Moreover, we show that this coherent transport follows different trajectories in opposite directions. Finally we present a scheme for finding the constants of motion in these configurations. It is important to note that even though we employ second quantization in our analysis, the investigated system is by no means quantum mechanical. In other words, we apply bosonic algebra only as a mathematical generator for higher order *classical* PT symmetric arrays.

We first start by considering a non-interacting PT symmetric two-site Bose-Hubbard Hamiltonian:

$$\hat{H}_2^{(1)} = -i\hbar\frac{\gamma}{2}\left(\hat{a}_1^+\hat{a}_1 - \hat{a}_2^+\hat{a}_2\right) + \hbar\kappa\left(\hat{a}_1^+\hat{a}_2 + \hat{a}_1\hat{a}_2^+\right) \qquad (1)$$

Here $\hat{a}_{1,2}^+$ and $\hat{a}_{1,2}$ are the bosonic creation and annihilation operators of site 1 and 2, respectively and they obey $\left[\hat{a}_i,\hat{a}_j\right] = \left[\hat{a}_i^+,\hat{a}_j^+\right] = 0$, $\left[\hat{a}_i,\hat{a}_j^+\right] = \delta_{ij}$ where $\delta_{ij}$ is the Kronecker delta function. The parameter $\gamma$ represents the gain/loss factor while $\kappa$ is the hopping constant between the two sites. Finally the superscript in $\hat{H}_2^{(1)}$ indicates that this Hamiltonian was obtained by quantizing the classical PT coupled waveguide/cavity strcutures (i.e. it is the first step in an iterative process that will be explained later) while the subscript denotes the number of sites The eigenvalue spectrum of Eq.(1) can be obtained in closed form by diagonalizing $\hat{H}_2^{(1)}$ through the transformations

$$\begin{pmatrix} \hat{b}_e \\ \hat{b}_o \end{pmatrix} = \begin{pmatrix} \cos(\alpha/2) & \sin(\alpha/2) \\ -\sin(\alpha/2) & \cos(\alpha/2) \end{pmatrix} \begin{pmatrix} \hat{a}_1 \\ \hat{a}_2 \end{pmatrix} \text{ with } \tan(\alpha) = \frac{2\kappa}{i\gamma}.$$ By doing so, we obtain

$$\lambda_{M,m} = (M-2m)\hbar\sqrt{\kappa^2 - (\gamma/2)^2}, \; m = 0,1,2,...,M \quad , \quad (2)$$

where $\lambda_{M,m}$ are the eigenvalues associated with stationary states having $m$ and $M-m$ bosons populating the even/odd-like modes, respectively. Note that the spectrum $\lambda_{M,m}$ has one exceptional point of order $M$ at $\gamma = 2\kappa$. Without any loss of generality, we consider the case of $M = 2N$ and employ the symmetrized bases

$$|n_{2N}\rangle = \frac{(\hat{a}_1^+)^{N-n}(\hat{a}_2^+)^{N+n}}{\sqrt{(N-n)!(N+n)!}}|vac\rangle = |N-n, N+n\rangle, \text{ with } N \mp n \text{ being the number of}$$

bosons in sites 1 and 2, respectively and $|vac\rangle$ is the vacuum state. In contrast to quantum systems, these bases are used here only to generate PT symmetric arrays and thus both integers and half integers are valid choices for $N$. The dynamical evolution of any arbitrary wavefunction inside this subspace is given by $|\psi(t)\rangle = \sum_{n=-N}^{N} c_n(t)|n_{2N}\rangle$ where the time dependent amplitudes $c_n(t)$ are completely determined by their initial conditions $c_n(0)$ and the coupled ordinary differential equations (ODEs):

$$i\hbar \frac{d\vec{c}}{dt} = \Omega\vec{c}, \quad \Omega = \begin{bmatrix} -i\gamma N & \kappa g_{-N+1} & & & \\ ... & ... & ... & ... & ... \\ & \kappa g_n & i\gamma n & \kappa g_{n+1} & \\ ... & ... & ... & ... & ... \\ & & & \kappa g_N & i\gamma N \end{bmatrix}, \quad (3)$$

where $\vec{c} = [c_{-N} \quad ... \quad c_{N-1} \quad c_N]^T$ with the superscript $T$ indicating matrix transpose. In Eq.(3), the coupling coefficients are symmetric around $n=0$ and are given by $g_n = \sqrt{(N+n)(N-n+1)}$. On the other hand, as expected the gain/loss profile (determined by the diagonal elements of $\Omega$) is antisymmetric and the system exhibit PT symmetry. Note that in the bases $|n_{2N}\rangle$, the eigenvalue spectrum of Eq.(3) is given by $\lambda_{2N,n} = -2n\hbar\sqrt{\kappa^2 - (\gamma/2)^2}$, $|n| \leq N$. These results were also obtained previously using angular momentum operator algebra [17,18]. Before we proceed, we would like to emphasize again that despite the quantum mechanical origin of the problem, Eq.(3) is a system of ODEs that can be simulated optically by classical light transport in coupled waveguide/cavity arrays. For example, Fig.1 (a) and (b) depict a schematic of waveguide/cavity arrays that can be used to emulate Eq.(3) when $N=1$ and $N=3/2$, respectively. The waveguides/cavities are represented by the nodes and the network connectivity (diagonal/off diagonal values of $\Omega$) is indicated in the figure. These structures can be realized using current photonic technology. For instance, the Hermitian analogue of the arrays in Fig.1 was experimentally investigated in optical waveguide lattices [16]. Moreover, PT symmetric optical waveguides, optical mesh periodic potentials, photonic arrays and microcavities were also experimentally investigated [11,12,19-21].

Next, we show that more complicated PT symmetric networks (with nonlinear topologies) can be generated by successive application of the aforementioned bosonic quantization procedure. As we will see, these structures exhibit an even richer spectral features than those generated from $\hat{H}_2^{(1)}$. We do so by replacing the $c$-numbers $c_n$ in Eq.(3) by a bosonic annihilation operators $\hat{a}_n$:

$$i\hbar \frac{d\vec{\hat{a}}}{dt} = \Omega \vec{\hat{a}} \quad , \quad (4)$$

where $\vec{a} = [\hat{a}_{-N} \ ... \ \hat{a}_n \ ... \ \hat{a}_N]^T$ and the formal solution of Eq.(4) is given by $\vec{\hat{a}}(t) = \exp(-i\Omega t)\vec{\hat{a}}(0)$. In order to proceed, we note that Eq.(4) is the Heisenberg equation of motion associated with the Hamiltonian:

$$\hat{H}_{2N+1}^{(2)} = i\gamma\hbar \sum_{n=-N}^{N} n a_n^+ a_n + \kappa\hbar \left( \sum_{n=-N+1}^{N} g_n \hat{a}_n^+ \hat{a}_{n-1} + \sum_{n=-N}^{N-1} g_{n+1} \hat{a}_n^+ \hat{a}_{n+1} \right) \quad (5)$$

Finally, expanding $\hat{H}_{2N+1}^{(2)}$ in its Fock space generates higher hierarchy PT symmetric structures. This recursive second quantization can be applied indefinitely to generate higher hierarchy of complex PT symmetric networks.

We elaborate on our results by considering a specific example. For instance, the second quantization of the PT symmetric configuration of Fig.1(a) is given by $\hat{H}_3^{(2)} = -i\gamma\hbar(\hat{a}_1^+\hat{a}_1 - \hat{a}_3^+\hat{a}_3) + \sqrt{2}\kappa\hbar(\hat{a}_1^+\hat{a}_2 + \hat{a}_1\hat{a}_2^+ + \hat{a}_2^+\hat{a}_3 + \hat{a}_2\hat{a}_3^+)$. As before, the Hamiltonian $\hat{H}_3^{(2)}$ can be also diagonalized by means of linear transformations and its eigenvalue spectrum is given by $\lambda_{M,m_1,m_2} = 2(M - 2m_1 - m_2)\hbar\sqrt{\kappa^2 - (\gamma/2)^2}$, $m_{1,2} = 0,1,2,...,M$ with the condition $m_1 + m_2 \leq M$. Here $m_{1,2}$ are the occupation numbers of the two lowest supermodes of $H_3^{(2)}$ while $M - m_1 - m_2$ represents the population of the third mode. The 2 and 3-boson representations of $\hat{H}_3^{(2)}$ are shown in Figs. 2(a) and (b), correspondingly. In contrast to those shown in Fig.1, we note that the PT symmetric networks depicted in Fig.2 exhibit nonlinear topology. Together with the structure of the Hamiltonian, this leads to several intriguing features in the eigenvalue spectrum. More specifically, the PT symmetric array depicted in Fig. 2(a) has two degenerate eigenmodes associated with the stationary states $|1,0,1\rangle_D$ and $|0,2,0\rangle_D$, where the subscript $D$ indicates a representation in the diagonal bases (where the numbers represent the occupation of the system's supermodes) as opposed to those defined by the site numbers. Moreover, these degenerate modes has null eigenvalue and they never undergo PT phase transition regardless of the values of $\gamma$. On the other hand, the network depicted in Fig.2 (b) exhibits even multiple sets of degenerate eigenstates. Similar to that of Fig.2(a), the supermodes corresponding to the states $|1,1,1\rangle_D$ and $|0,3,0\rangle_D$ have zero eigenvalue and they never experience PT phase transition. On the other hand, each of the two sets of eigenstates $\{|0,2,1\rangle_D, |1,0,2\rangle_D\}$ and $\{|1,2,0\rangle_D, |2,0,1\rangle_D\}$ is doubly degenerate and their eigenvalues are complex conjugate. Higher order symmetries can be also found in higher hierarchy networks.

Figure 3 illustrates the intensity distribution (or probability amplitudes) $|c_n|^2$ associated with the degenerate eigenmodes $|1,0,2\rangle_D$ and $|0,2,1\rangle_D$ of the structure in Fig.2 (b). More specifically, the eigensolutions associated with the PT symmetric phase ($\gamma < 2\kappa$) are shown in Fig.3 (a) and (b) while the broken phase scenario is depicted in (c) and (d). It is remarkable that these degeneracies persist even when the PT symmetry is spontaneously broken. This fact is manifested in Figs.3 (c) and (d) where two different eigenmodes share the same complex eigenvalues and hence experience the same amplifications during evolution.

We note that these degeneracies do not arise from geometric transformations (rotation, reflection, etc) that leave the network topology invariant but are rather an outcome of hidden symmetries. The nature of these degeneracies can be characterized by a set of operators $\{\hat{A}_i\}$ where $[\hat{H}, \hat{A}_i] = 0$ and $[\hat{A}_i, \hat{A}_j] \neq 0$ [22]. In our system, these operators can be easily identified in the diagonal bases. For example, the diagonal form of $\hat{H}_3^{(2)}$ can be written as $\hat{H}_{D,3}^{(2)} = -\hbar\sqrt{4\kappa^2 - \gamma^2}\left(\hat{q}_1^+ \hat{q}_1 - \hat{q}_3^+ \hat{q}_3\right)$, where $\hat{q}_i^+$ and $\hat{q}_i$ are the creation and annihilation operators associated with the supermodes and as before, the subscript $D$ denotes a diagonal representation. It is now straightforward to show that the operators $\hat{A}_1 = \hat{q}_1^+ \hat{q}_2 \hat{q}_2 \hat{q}_3^+$ and $\hat{A}_2 = \hat{q}_1 \hat{q}_2^+ \hat{q}_2^+ \hat{q}_3$ satisfy the above relations and thus give rise to degeneracy in the spectrum. For example, $\hat{A}_1 |0,3,0\rangle_D = |1,1,1\rangle_D$ and $\hat{A}_2 |1,1,1\rangle_D = |0,3,0\rangle_D$ while $\hat{A}_1 |0,2,1\rangle_D = |1,0,2\rangle_D$ and $\hat{A}_2 |1,0,2\rangle_D = |0,2,1\rangle_D$. It is straightforward to see that similar mathematical analysis applies for higher order Hamiltonians generated using more iteration. This can be best illustrated by investigating the origin of these degeneracies from a more intuitive perspective. Figure 4 depicts the bosons occupation numbers for three pairs of degenerate states associated with the Hamiltonian $\hat{H}_{D,3}^{(2)}$ where the eigenvalue levels (analogous to energy levels in atoms) are shown by yellow thick lines while individual bosons are represented using red spheres. Evidently, these degeneracies arise from the existence of different possible boson distribution profiles that share the same eigenvalue. It is now clear that higher order

Hamiltonians that exhibiting more than three 'energy' levels will even have more bosonic distribution configurations that share the same eigenvalues and thus more degeneracies are expected for networks generated by using higher iterations.

This existence of these degeneracies is an intriguing feature that merits further investigations. For example, it would be of interest to investigate the interplay between nonlinear interactions and hidden symmetries on the lasing characteristics of multi-cavity photonic molecules having similar topologies. We carry these investigations somewhere else.

Next we investigate light transport in these PT symmetric photonic structures and we study the associated conservation laws. Our numerical analysis indicates that perfect state transfer is possible in these geometries only for certain initial conditions. In order to demonstrate this feature, we plot the evolution of optical field intensities for the configuration shown in Fig.2 (a) in the PT symmetric phase under two different excitations. Figure 5(a) shows light transport dynamics under the initial condition $c_n(0) = \delta_{n,1}$. Perfect state transfer from $c_1$ to $c_6$ (waveguide numbering scheme is depicted in the figure) occurs after a propagation distance $l_1$. Fig.5(b) shows the full revival cycle associated with Fig.5(a). Note that coherent transport in the opposite direction (from $c_6$ to $c_1$) occurs after a distance $l_2 \neq l_1$. Moreover, the optical power levels depend on the direction of transport (left to right versus right to left). Finally, Fig.5 (c) depicts the same quantities when $c_n(0) = \delta_{n,2}$. Clearly no perfect transfer is observed in this case. This conditional PST in non-Hermitian lattices was also highlighted in [18] in the special case when the propagation distance corresponds to the revival half cycle. In a more general scenario, coherent transport length in non-Hermitian lattices is not necessarily restricted to the evolution half cycle $(l_1 + l_2)/2$.

We now explore the conservation laws associated with light transport in these PT-invariant networks. The simple case of two coupled PT symmetric waveguides were treated using Stokes parameters [15]. However, these calculations become cumbersome for complicated configurations. Here we present an alternative route for obtaining the system's constants of motion using standard matrix algebra. In quantum mechanics, conservation laws arise from symmetry operators that commute with the Hermitian Hamiltonian. However, for any general non-Hermitian Hamiltonian $\hat{H}$, the conserved quantities $Q_i$ are associated with operators $\hat{S}_i$ that satisfy the relation $\hat{S}_i\hat{H} - \hat{H}^+\hat{S}_i = 0$; and are given by $Q_i = \langle \hat{S}_i \rangle$. For discrete systems, the above operators reduce to matrices and the conserved quantities can be obtained by solving the above modified commutation relation for the unknown elements of the matrix $S_i$. To illustrate the expediency of this technique, we consider the PT symmetric network described by Eq.(3) when $\gamma \neq 0$ and we treat the case of $N = 1$ (Fig.1 (a)). In this case, the above procedure yields the condition $s_{11} = s_{33}$, where $s_{ij}$ represents that elements of matrix $S$. Moreover we find that the linear system of equations is of rank five and we choose $s_{12}$, $s_{22}$ and $s_{32}$ as free parameters. We then obtain the independent solutions by identifying three different vectors that span the parameter space $(s_{12}, s_{22}, s_{32})$. One particular choice of these bases gives:

$$S_1 = \begin{bmatrix} 0 & 0 & 1 \\ 0 & 1 & 0 \\ 1 & 0 & 0 \end{bmatrix}, \quad S_2 = \begin{bmatrix} 0 & 1 & i\gamma/\sqrt{2}\kappa \\ 1 & 0 & 1 \\ -i\gamma/\sqrt{2}\kappa & 1 & 0 \end{bmatrix}, \quad S_3 = \begin{bmatrix} 1 & i\gamma/\sqrt{2}\kappa & -(\gamma^2/2\kappa^2+1) \\ -i\gamma/\sqrt{2}\kappa & 0 & i\gamma/\sqrt{2}\kappa \\ -(\gamma^2/2\kappa^2+1) & -i\gamma/\sqrt{2}\kappa & 1 \end{bmatrix}$$

and the associated conserved quantities $Q_{1,2,3} = \vec{c}^+ S_{1,2,3} \vec{c}$ are $Q_1 = (c_1^* c_3 + c_1 c_3^*) + |c_2|^2$,

$$Q_2 = (c_1^* c_2 + c_1 c_2^*) + (c_3^* c_2 + c_3 c_2^*) + \frac{i\gamma}{\sqrt{2}\kappa}(c_1^* c_3 - c_1 c_3^*) \qquad \text{and}$$

$$Q_3 = |c_1|^2 + |c_3|^2 + \frac{i\gamma}{\sqrt{2}\kappa}(c_1^* c_2 - c_1 c_2^*) - \frac{i\gamma}{\sqrt{2}\kappa}(c_3^* c_2 - c_3 c_2^*) - \left(\frac{\gamma^2}{2\kappa^2}+1\right)(c_1^* c_3 + c_1 c_3^*).$$

Evidently, the above calculation is straightforward and can be generalized to any PT-symmetric network of any complexity and dimensionality.

In conclusion, we have introduced a recursive bosonic quantization technique for generating classical PT photonic structures that possess hidden symmetries and higher order exceptional points. We have also investigated the nature of the eigensolutions as well as light transport in these geometries and we have demonstrated that perfect state transfer is possible only for certain initial conditions. Moreover, we have shown that this coherent transport is asymmetric with field amplitudes following two different trajectories in opposite transport directions. A general scheme for identifying the conservation laws in such PT-symmetric photonic networks has been also presented.

**Figure captions**

Figure 1: (Color online) (a) and (b) depict a schematic of waveguide/cavity arrays that can be used to emulate Eq.(3) when $N=1$ and $N=3/2$, respectively. The waveguides/cavities are represented by the nodes and the network connectivity (diagonal/off diagonal values of $\Omega$) is indicated in the figure. Distances between elements are not shown to scale.

Figure 2: (Color online) A schematic of the coupled photonic structures that correspond to 2 and 3-boson representations of $\hat{H}_3^{(2)}$ are shown (a) and (b), respectively. Similar to Fig.1, the diagonal and off diagonal elements are also indicated in the figure. Distances between elements are not shown to scale.

Figure 3: (Color online) (a) and (b) depict the intensity distribution (or probability amplitudes) $|c_n|^2$ associated with the degenerate eigenmodes $|1,0,2\rangle_D$ and $|0,2,1\rangle_D$ when $\kappa=1$ and $\gamma=1$ while (c) and (d) illustrate the same modes when $\kappa=1$ and $\gamma=3$. Note that the onset of PT phase transition is $\gamma=2\kappa$. The transparent cylinders represent the waveguide structure.

Figure 4: (Color online) Depicts bosons occupation numbers of some degenerate eigenstates (each degenerate pair is schematically shown in one row) associated with the Hamiltonian $\hat{H}_{D,3}^{(2)} = -\varepsilon\left(\hat{q}_1^+ q_1 - \hat{q}_3^+ q_3\right)$, where $\varepsilon = \hbar\sqrt{4\kappa^2 - \gamma^2}$. The eigenvalue levels

(analogous to energy levels in atoms) are shown by yellow thick lines while individual bosons are represented using red spheres. For instance, it is clear that bosonic occupation configurations shown in the first row have null eigenvalue. The action of hidden symmetry operators $A_{1,2}$ is also indicated in the figure. Evidently even higher order degeneracies will arise for Hamiltonians associated with more than three sites.

Figure 5(a) (Color online) (a) Light transport in the array shown in Fig. 2(a) under the initial excitations: $c_n(0) = \delta_{n,1}$. Note that perfect state transfer from $c_1$ to $c_6$ occurs after a propagation distance $l_1$. (b) Depicts the revival dynamics after one full cycle where coherent transport in the opposite direction (from $c_6$ to $c_1$) occurs after $l_1 \neq l_2$. Note that the optical power levels depend on the transport direction. (c) Demonstrates the absence of PST when the initial excitation is $c_n(0) = \delta_{n,2}$. The transparent cylinders in (a) represent the individual waveguides. The array numbering scheme is also indicated in (a).

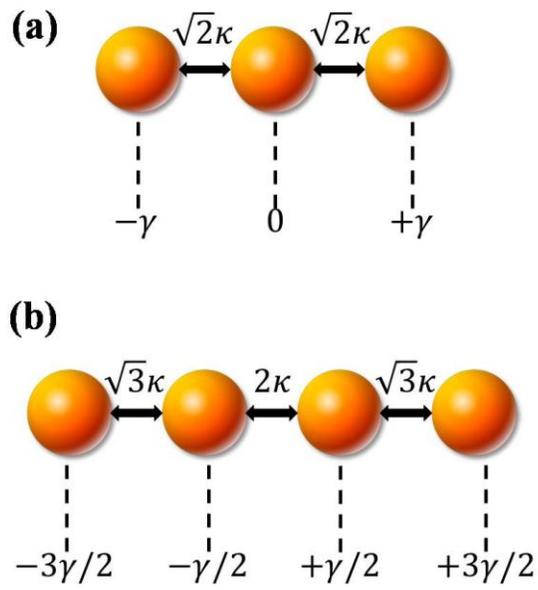

Fig.1

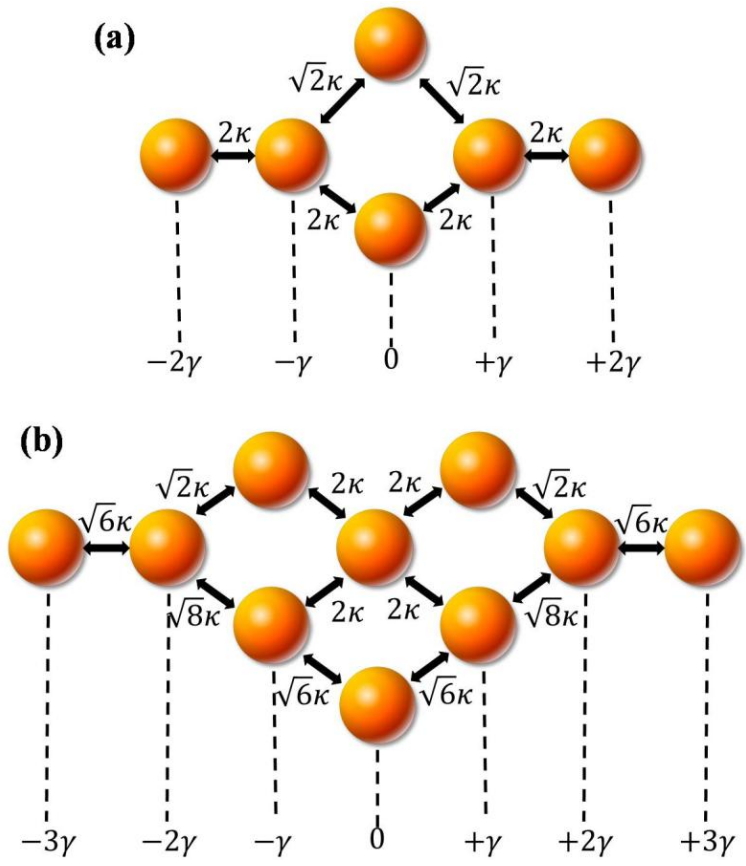

Fig.2

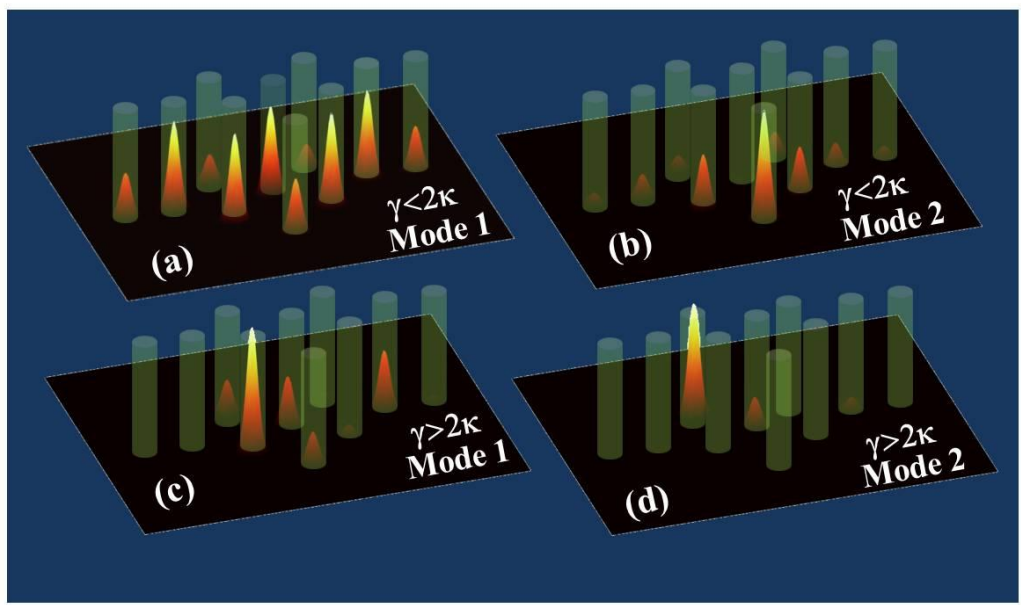

Fig.3

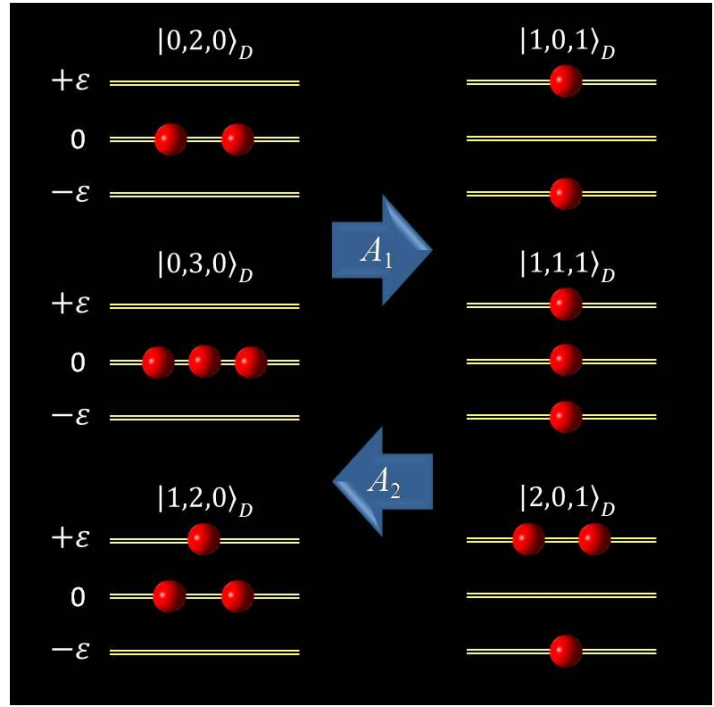

Fig.4

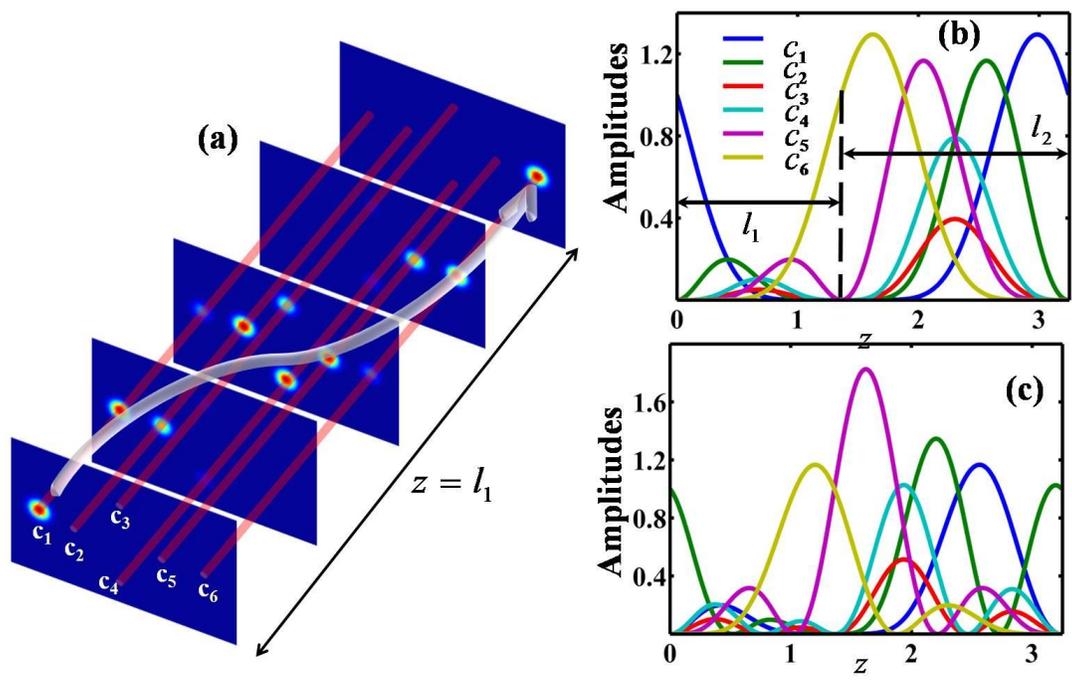

Fig.5